\documentclass[aps,prl,groupedaddress,showpacs,twocolumn]{revtex4-1}

\usepackage{amsmath}
\usepackage{graphicx}
\usepackage{dcolumn}
\usepackage{bm}
\usepackage{color}
\usepackage{soul}

\begin{document}

\title{
Heteroclinic structure of parametric resonance in the nonlinear Schr\"odinger equation\\
}

\author{M. Conforti$^{1\dagger}$, A. Mussot$^{1}$, A. Kudlinski$^{1}$, S. Rota Nodari$^{2}$, G. Dujardin$^{3}$, S. De Bi\'evre$^{3}$, A. Armaroli,$^4$ and S. Trillo$^{5*}$}

\affiliation{
$^1$Univ. Lille, CNRS, UMR 8523 - PhLAM - Physique des Lasers Atomes et
Mol\'ecules, F-59000 Lille, France\\
$^2$IMB UMR 5584, CNRS, Univ. Bourgogne Franche-Comt\'e, F-21000 Dijon, France\\
$^3$Univ. Lille, CNRS, UMR 8524 - Laboratoire Paul Painlev\'e; Equipe MEPHYSTO, 
INRIA, F-59000 Lille,
France\\
$^4$FOTON (CNRS-UMR 6082) Universit\'e de Rennes 1, ENSSAT, 6 rue de Kerampont, CS 80518, 22305 Lannion Cedex, France\\
$^5$Dipartimento di Ingegneria, Universit\`a di Ferrara, Via Saragat 1, 44122 Ferrara, Italy
}

\email{stefano.trillo@unife.it} 
\email[$^\dagger$]{matteo.conforti@univ-lille1.fr} 

\date{\today} 
 
\begin{abstract} 
We show that the nonlinear stage of modulational instability induced by parametric driving in the {\em defocusing} nonlinear Schr\"odinger equation can be accurately described by combining mode truncation and averaging methods,
valid in the strong driving regime. The resulting integrable oscillator reveals a complex hidden heteroclinic structure of the instability. 
A remarkable consequence, validated by the numerical integration of the original model, is the existence of breather solutions separating different Fermi-Pasta-Ulam recurrent regimes.
Our theory also shows that optimal parametric amplification unexpectedly occurs outside the bandwidth of the resonance (or Arnold tongues) arising from the linearised Floquet analysis.

\end{abstract} 
 
 
\maketitle
\begin{figure}[h]
\centering
\includegraphics[width=\columnwidth]{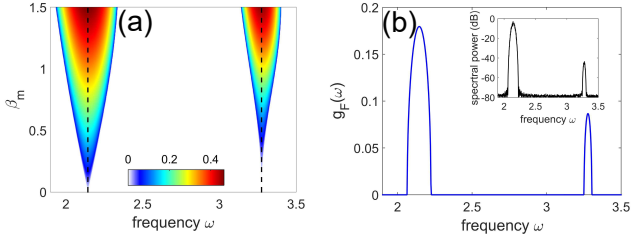}
     \caption{(Color online) Results of the linear Floquet analysis for $f_{\Lambda}(z)=\cos(k_g z)$, $\Lambda=1$, $P=1$: (a) false color plot showing first two MI tongues in the plane $(\omega, \beta_m)$ [dashed vertical lines stand for $\omega_p$, $p=1,2$, from Eq. (\ref{omega_m})]; (b) section at $\beta_m=0.5$ showing gain curves $g_F(\omega)$; Inset: spectral output ($z=50$) from NLSE (\ref{nls}) numerical integration, where PRs grow out of white noise superimposed onto $\psi_0$.}
     \label{f1}
\end{figure}

 
Following the pioneering studies by Faraday and Lord Rayleigh \cite{history}, the universal nature of parametric resonances (PRs) induced by periodic variations of a system parameter \cite{NMbook} was established in applications involving standard \cite{pendulum}, nano- \cite{Turner98} and micro-oscillators \cite{NMEMs}, optical trapping \cite{DiLeonardo07}, 
control of solitons \cite{discrete}, 
wrinkling \cite{wrinkling}, and ship motion \cite{ships}, or in fundamental studies of universe inflation \cite{inflation}, quantum tunnelling \cite{quantumtunneling}, and Faraday waves \cite{Faraday1,Faraday2}. 
While the concept of PR originates in the linear world \cite{mathieu}, 
PRs deeply impact also the behavior of {\em nonlinear} conservative systems.
However, the full nonlinear dynamics of PRs is relatively well understood only for low-dimensional Hamiltonian systems \cite{NMbook,NLMathieu,Verhulst}. Conversely, the analysis of {\em extended} systems described by PDEs  with periodicity {\em in the evolution variable} \cite{transverseperiodic} is essentially limited to determine the region of parametric instability (Arnold tongues) via Floquet analysis \cite{thDOF,Garciaprl99,Rapti04,Centurion06}, while the nonlinear stage of PR past the linearized growth of the unstable modes remains mostly unexplored.  
\newline \indent
In this letter, taking the periodic defocusing nonlinear Schr\"odinger equation (NLSE) as a widespread example describing, e.g. periodic management of atom condensates \cite{Faraday1,Garciaprl99,Rapti04,BECmanage}, optical beam propagation in layered media \cite{Centurion06}, or optical fibers with periodic dispersion \cite{thDOF,Armaroli12,expDOF}, we show that the PR gives rise to quasi-periodic evolutions which exhibit on average Fermi-Pasta-Ulam (FPU) recurrence \cite{FPUrec} with a remarkably complex (but ordered) underlying phase-plane structure. 
Such structure describes the continuation into the nonlinear regime of the modulational instability (MI) of a background solution, uniquely due to the parametric forcing (with zero forcing the defocusing NLSE is stable). A byproduct of this structure is the existence of breather-like solutions \cite{Akh85}. This fact suggests the intriguing possibility of observing rogue waves \cite{rogue}, which are commonly associated with breathers \cite{rogue2}, in the defocusing NLSE \cite{baronio}. On the other hand, the richness of such structure allows us to remarkably predict that optimal parametric amplification occurs at a critical frequency where the system lies off-resonance (outside the PR bandwidth). 
Our approach retains its validity in the most interesting regime of {\em strong} parametric driving, where the system is found to exhibit a remarkably ordered structure despite its broken translational symmetry and integrability. In this sense the physics differs from other integrable models exhibiting a complex nonlinear dynamics of MI already in the undriven regime (e.g. focusing NLSE \cite{Akh85,Ablowitz89,Moon90,TW91,ErkintaloPRL11,ZG13,Chabchoub14,Biondini15}), around which chaos can develop under weak periodic perturbations \cite{Moon90,Ercolani90,Forest92}.\\
\indent We consider the following periodic NLSE
\begin{equation} 
i\frac{\partial \psi}{\partial z} - \frac{\beta(z)}{2} \frac{\partial^2 \psi}{\partial t^2} + |\psi|^2 \psi=0,\label{nls} 
\end{equation}
referring, without loss of generality, to the notation used in optical fibers in suitable scaled units.  The dispersion is $\beta(z) = \beta_{av} + \beta_m f_{\Lambda}(z)$, with positive average $\beta_{av}>0$ (equivalent to the defocusing regime); $f_{\Lambda}(z)$ has period $\Lambda=2\pi/k_g$, zero mean and minimum $-1$. The method can be easily extended to deal also with periodic nonlinearities. We are interested in the nonlinear evolution of perturbations of the stationary background solution $\psi_0=\sqrt{P} \exp(i P z)$ with power $P=|\psi_0|^2$ \cite{norm}.
First, we briefly recall the origin of PRs in this system \cite{thDOF,Armaroli12,expDOF}. MI of $\psi_0$ can be analyzed by inserting in Eq. (\ref{nls}) the ansatz $\psi=\psi_0+ a(z,t)$, $a$ being a perturbation at frequency $\omega$ of the general form $a=[\epsilon_1(z) \exp(i\omega t) + \epsilon_2^*(z) \exp(-i\omega t)]\exp(i P z)$. Linearizing around ${\bf x}(z)=[\epsilon_1(z) ,~ \epsilon_2(z)]^T$ gives a $\Lambda$-periodic problem that can be treated 
by means of Floquet theory \cite{Armaroli12,thDOF}. In the absence of perturbation ($\beta_m=0$), ${\bf x}(z)$ exhibits only phase changes ruled by imaginary eigenvalues $\pm i k$, where $k^2=\beta_{av}\omega^2/2 \left(\beta_{av}\omega^2/2 + 2 P \right)$ represents the squared spatial frequency of the evolution. As a result $\psi_0$ is stable. Any arbitrarily small perturbation $\beta_m \neq 0$ induces, regardless of its shape $f_{\Lambda}(z)$, instability at multiple frequencies ($p=1,2,\ldots$)
\begin{eqnarray}\label{omega_m}
\omega_p=\sqrt{ \frac{2}{\beta_{av}}  \left( \sqrt{ P^2+\left( \frac{p \pi}{\Lambda} \right)^2  } - P \right) },
\end{eqnarray}
which fulfil the PR condition $k_g=2k(\omega_p)/p$, analogous to Mathieu equation (though for spatial frequencies, instead of temporal ones).
The Floquet analysis gives rise to instability islands, or Arnold tongues, in the plane $(\omega, \beta_m)$, with $\omega_p$ representing the tip of the tongues, as shown in Fig. \ref{f1}(a), taking as an example $f_{\Lambda}(z)=\cos(k_g z)$ and $p=1,2$. Figure \ref{f1}(b) shows the instability gain spectrum $g_F(\omega)$ at $\beta_m=0.5$, which accurately predicts the spontaneous growth of MI bands from white noise, obtained from NLSE integration [inset in Fig. \ref{f1}(b)].

Two aspects of the PR instability are of crucial importance: (i) it exhibits narrowband features around the tongue tip frequencies $\omega_p$; 
(ii) different $\omega_p$ are generally incommensurate, which greatly reduces the possibility that the harmonics of a probed frequency experience exponential amplification due to higher-order bands. Under such circumstances, three-mode truncations constitute a suitable approach to describe the underlying structure of the dynamics \cite{Whitham,Bishop90,TW91,Millot98}. However, unlike uniform media where the truncation is integrable, in the PR such structure can remain hidden owing to the fast phase variations induced by the parametric driving, which breaks the integrability of the truncated model too. In order to unveil the dynamics, we need to combine the mode truncation approach with suitable phase transformations and averaging \cite{Clausen97}. We start by substituting in Eq. (\ref{nls}) the field $\psi=A_0(z) + a_1(z) \exp(-i \omega t) + a_{-1}(z) \exp(i \omega t)$ and group all nonlinear terms vibrating at frequencies $0, \pm \omega$, neglecting higher-order harmonic generation. For sake of simplicity we consider henceforth the case of symmetric sidebands $a_1 = a_{-1} \equiv  A_1/\sqrt{2}$, though our analysis and conclusions straightforwardly extend to the case $a_1 \neq a_{-1}$. We obtain the following non-autonomous Hamiltonian system of ODEs (dot stands for $d/dz$) 
\begin{eqnarray}
-i \dot{A_0} &=& (|A_0|^2+2|A_1|^2) A_0 + A_1^2 A_0^*,\label{tr1} \\
-i \dot{A_1} &=& 
\!\left[  {\textstyle \frac{ \beta(z)\omega^2}{2}}+ \left( {\textstyle \frac{3|A_1|^2}{2}}+2|A_0|^2 \right) \right]\!A_1 + A_0^2 A_{1}^*,\label{tr2}
\end{eqnarray} 
where the only conserved quantity, i.e. $P=|A_0|^2 + |A_1|^2$, is not sufficient to guarantee integrability. 
In order to describe the mode mixing in the $p$-th unstable PR band beyond the linearized stage, we transform to new phase-shifted variables $u(z), w(z)$, defined as
\begin{equation}
A_0(z)=u(z); \, A_1(z)=w(z) e^{i \left( p \frac{k_g}{2}z + \frac{\delta k(z)}{2} \right)},
\end{equation}
where $\delta k(z)= \beta_m \omega^2 \int_0^z f_{\Lambda}(z') dz'$ physically accounts for the oscillating wavenumber mismatch of the three-wave interaction.
Then, we exploit the general Fourier expansion $\exp[i \delta k(z)] = \sum_n c_n \exp(-i n k_g z)$, which allows us to cast Eqs. (\ref{tr1}-\ref{tr2}) in the form
\begin{flalign}
-i \dot{u} &=(P + |w|^2)u + \left[ c_p + F_{\Lambda}(z) \right] w^2 u^*,\label{uw1} \\
-i \dot{w} &=\left( \frac{\kappa}{2} + \frac{|w|^2}{2}  + |u|^2 \right) w  + \left[ c_p^* + F^*_{\Lambda}(z) \right] u^2 w^*,\label{uw2}
\end{flalign}
where $F_{\Lambda}(z) \equiv \sum_{n \ne p}c_n \exp[-i(n-p) k_g z]$, and $\kappa \equiv \beta_{av} \omega^2 - p k_g+ 2 P$ measures the mismatch from optimal linearized amplification. 
 Indeed  $\kappa =0$ is equivalent to the quasi-phase-matching condition $\beta_{av} \omega^2 + 2P = p k_g$, where the quasi-momentum $p k_g$ associated to the forcing compensates for the average nonlinear wavenumber mismatch of the three-wave interaction \cite{expDOF}. In the quasi-matched regime ($|\kappa| \ll 1$), the dominant mixing terms $c_p w^2 u^*$ and $c_p^*u^2 w^*$ in Eqs. (\ref{uw1}-\ref{uw2}) are responsible for the growth of sidebands associated with the PR instability in the $p$-th band. However, additional contributions to the mixing arise from the mismatched terms contained in the $\Lambda$-periodic function $F_{\Lambda}(z)$. In order to evaluate their impact we generalize the approach of Ref. \cite{Clausen97} developed for quadratic media. 
We assume $1/k_g$ to be small and expand $u,w$ in Fourier series $u(z) = \sum_n u_n(z)e^{inpk_gz}$, $w(z)=\sum_n w_n(z)e^{inpk_gz}$. Assuming that $w_n,u_n$ vary slowly with respect to $\exp(ik_gz)$, and the harmonics to be of order $1/k_g$ (or smaller, compared to leading order or spatial average $u_0, w_0$), we are able to express $w_n,u_n$  through the  relations $u_n=\frac{1}{pk_g}\frac{c_{p(1-n)}}{n}w_0^2u_0^*$, $w_n=\frac{1}{pk_g}\frac{c^*_{p(1+n)}}{n}u_0^2w_0^*$,
which allow us to obtain a self-consistent system for $u_0(z), w_0(z)$
%
\begin{align}
\nonumber -i \dot{u_0} =&(P+|w_0|^2)u_0 + c_p w_0^2 u_0^*+ \\
&\qquad \qquad \qquad \quad +\alpha\left(|w_0|^4-2|w_0 u_0|^2\right)u_0,\label{av1}\\
-i \dot{w_0} =& \left( \frac{\kappa}{2} + \frac{|w_0|^2}{2}  + |u_0|^2 \right) w_0 + c_p^* u_0^2 w_0^*+\qquad \qquad \nonumber \\
&\qquad \qquad \qquad \quad -\alpha \left(|u_0|^4-2|w_0 u_0|^2\right)w_0, \label{av2} 
\end{align}
which shows that the mismatched terms result into an effective quintic correction weighted by the (small) coefficient $\alpha=\frac{1}{p k_g}\sum_{n\ne 0}\frac{|c_{p(1-n)}|^2}{n}$.
Equations (\ref{av1}-\ref{av2}) can be cast in Hamiltonian form
\begin{eqnarray} 
\dot{\eta} &=& -\frac{\partial H_p}{\partial \phi};
\;\;\dot{\phi} =\frac{\partial H_p}{\partial \eta} \label{Hav}, \\
H_p  &=& |c_p| \eta (1-\eta) \cos 2\phi + \frac{\kappa}{2}\eta -\frac{3}{4}\eta^2-\alpha \eta \left(1-3\eta+2\eta^2 \right),\nonumber 
\end{eqnarray}
in terms of fractional sideband power $\eta=|w_0|^2\approx|A_1|^2$ and overall phase $\phi=Arg[w_0(z)] - Arg[u_0(z)] + \phi_p/2$, $\phi_p=Arg[c_p]$.

\begin{figure}[h]
\centering
\includegraphics[width=7cm]{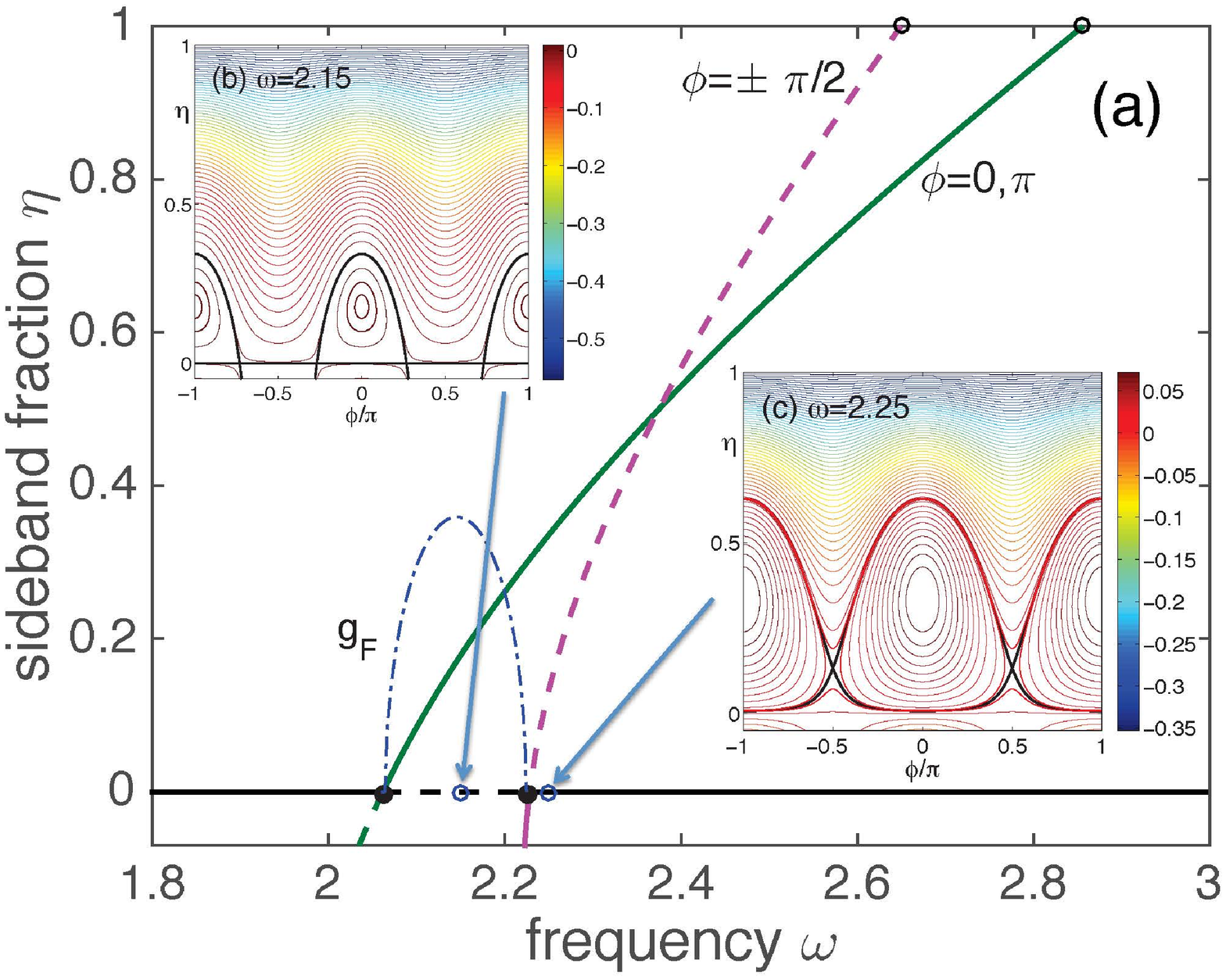} 
     \caption{(Color online) (a) Bifurcation diagram from Eqs. (\ref{Hav}): sideband fraction $\eta$ of unstable (dashed red) and stable (solid green) branches vs. $\omega$. The instability range of the pump mode $\eta=0$ coincides with the bandwidth calculated from Floquet analysis (gain $g_F$, dot-dashed line). Insets (b,c): phase-plane pictures for (b) $\omega=2.15$ inside PR gain bandwidth; (c) $\omega=2.25$, outside PR gain bandwidth, where the topology is affected by saddle eigenmodulations with $\eta \neq 0$. Here $p=1$ (primary PR), $\beta_m=0.5$, $\Lambda=1$, $P=1$. }
     \label{f2}
\end{figure}

Equations (\ref{Hav}) constitute an averaged integrable description of the fully nonlinear stage of the instability, which holds valid regardless of the choice of order $p$ and the specific function $f_{\Lambda}(z)$  \cite{transform}. Among the different tests that we have performed, in the following we present the results obtained for the harmonic case $f_{\Lambda}(z)=\cos(k_g z)$ already considered in Fig. \ref{f1}.
In this case, the Fourier expansion  $\exp[i \delta k(z)] =   \sum_{n=-\infty}^{\infty}(-1)^n J_{n}\left( \case{\beta_m \omega^2}{k_g} \right)\exp(-ink_g z)$ gives for the primary PR ($p=1$), $c_1(\omega)=-J_{1}\left( \frac{\beta_m \omega^2}{k_g} \right)$ (hence  $\phi_1=\pi$), and a safe approximation for the quintic correction is $\alpha (\omega) \approx \case{1}{k_g}\left(|c_0|^2 + |c_1|^2/2 \right)$, where $c_0(\omega)=J_0 \left( \case{\beta_m \omega^2}{k_g} \right)$, since $c_0 \gg c_{n}, n>1$.

\begin{figure}[htp]
\centering 
\includegraphics[width=\columnwidth]{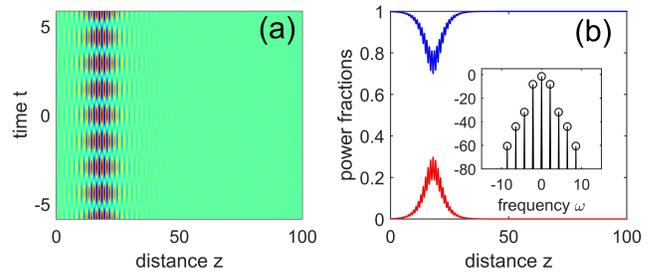}
     \caption{(Color online) PR breather excitation from numerical integration of NLSE (\ref{nls}): (a) color level plot of $|\psi|^2$; (b) fractions $|A_0|^2$, $|A_1|^2$ of Fourier modes vs. $z$. Inset: log scale spectrum at the point of maximum depletion, $z=18$. Here $\beta_m=0.5$, $\omega=2.15$, $\Lambda=1$, $P=1$, and initial condition $\eta_0=0.001$, $\phi_0=0.24162\pi$ corresponds to the separatrix in Fig. 2(b).}
     \label{f3}
\end{figure}

Explicit solutions of Eqs. (\ref{Hav}) can be written in terms of hyperelliptic functions. However, their phase-plane representation (level set of $H_p$) along with the bifurcation analysis are sufficient to gain a full physical insight. Figure \ref{f2} shows the bifurcation diagram, i.e. the value $\eta$ of the stationary points (solutions of $\dot{\eta}=\dot{\phi}=0$) versus frequency $\omega$. The instability of the pump mode $\eta=0$ reflects the PR instability of order $p$. Indeed $\eta=0$, $\phi^{\pm}=\pm \case{1}{2} \cos^{-1}[(\alpha-\kappa/2)/|c_p|]$ turn out to be saddle points of the Hamiltonian $H_p$ in the range of frequencies implicitly determined by the condition $ - |c_p(\omega)| \le \alpha(\omega)-\kappa(\omega)/2 \le |c_p(\omega)|$, which agree with the PR bandwidth from linear Floquet analysis [see the comparison in Fig. \ref{f2} for $p=1$]. Within such range of frequencies, the accessible portion of the phase plane ($\eta \ge 0$) is characterized by a heteroclinic separatrix which connects such saddles, dividing the phase plane into regions of inner and outer orbits which are similar to those describing librations and rotations of a standard pendulum, respectively [see Fig. \ref{f2}(b)]. At the edges of such frequency span, the pump mode bifurcates and new phase-locked eigenmodulation branches appear with modulation depth $\eta=\eta_s \neq 0$ variable with frequency, and phase locked either to $\phi=0,\pi$ (stable, centers) or $\phi=\pm \pi/2$ (unstable, saddles).  
New heteroclinic connections emanate from the latter, dividing the accessible phase plane into three different domains [see Fig. \ref{f2}(c)].

The structure illustrated in Fig. \ref{f2} has deep implications for the long-term evolution of the PR in the full NLSE (\ref{nls}). In order to show this we numerically integrate Eq. (\ref{nls}) with initial value representing a weakly modulated background: $\psi_0(t)=\sqrt{1-\eta_0}+\sqrt{2\eta_0} \exp(i \theta_0) \cos(\omega t)$, $\eta_0 \ll 1$, where $\theta_0$ is linked to the overall initial phase $\phi_0=\phi(0)$ as $\phi_0=\theta_0+\phi_p/2$. 


Considering first frequencies within a PR band, we show in Fig. \ref{f3} the excitation of the infinite-dimensional analog of the heteroclinic separatrix shown in the left inset in Fig. \ref{f2}, obtained from a very weak modulation ($\eta_0=0.001$) with suitable phase. This entails a single cycle of amplification connecting the background to itself with opposite phases, i.e. the analog of the well known Akhmediev breather 
of the integrable {\em focusing} NLSE \cite{Akh85} (see also \cite{Ablowitz89,Moon90,TW91,ErkintaloPRL11,ZG13,Chabchoub14}). This type of solutions of the periodic NLSE (\ref{nls}), which we term  as parametric resonance breathers (PR breathers), are characterized by a main breathing occurring on top of the short $\Lambda$-scale breathing. PR breathers can be excited for all frequencies inside the PR bandwith. We remark that although they entail the generation of harmonics of the input modulation, 
the spectrum decays rapidly as shown in the inset of Fig. \ref{f3}(b) and the dynamics is dictated by the first sideband pair with the harmonics that remain locked to them.

\begin{figure}[t]
\centering
\includegraphics[width=1\columnwidth]{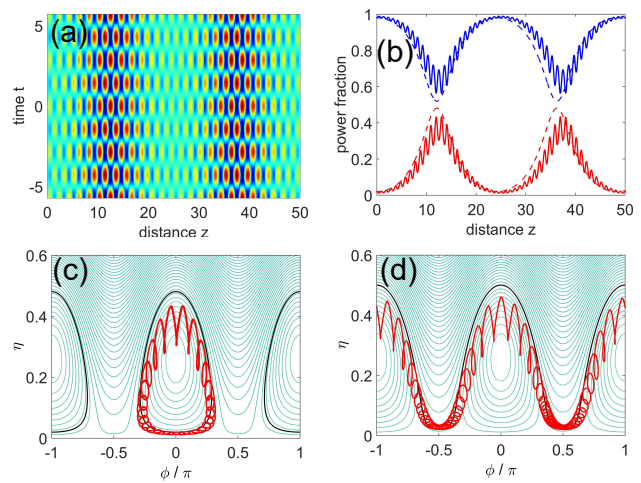} 
     \caption{(Color online) Quasi-periodic recurrent evolution from full NLSE numerical integration with $\eta_0=0.02$: (a) colormap of $|\psi|^2$; (b) evolution of extracted pump and sideband power fractions for $\phi_0=0$  (solid lines), compared with those from the average model (dashed lines), Eq. (\ref{Hav}). 
     (c-d) projections of the NLSE numerical evolutions over the phase plane of the averaged system for $\phi_0=0$ (c) and $\phi_0=\pi/2$ (d). Here $\beta_m=0.5$, $\omega=2.2$, $\Lambda=1$, $P=1$. }
     \label{f4}
\end{figure}

A PR breather divides the phase-plane into two types of dynamical behaviors which exhibit different FPU-like recurrence, i.e. cyclic amplification and de-amplification of the modulation over scales much longer than the $\Lambda-$scale of small oscillations. One of such recurrent regimes is displayed in Fig. \ref{f4}(a,b), obtained for $\eta_0=0.02$, $\phi_0=0$. When we flip the initial phase to $\phi_0=\pi/2$ we observe a very similar behavior (not shown).
However the projection of the NLSE evolutions onto the phase-space $(\eta,\phi)$ reveals very different behaviors for the two initial phases. While in both cases we observe quasi-periodic evolutions, in one case ($\phi_0=0$) the recurrence occurs around the libration-type of orbit of the averaged system [Fig. \ref{f4}(c)], whereas the recurrent dynamics for $\phi_0=\pi/2$ follows rotation-type of dynamics with the phase spanning continuously the full range $(-\pi,\pi)$ [Fig. \ref{f4}(d)]. This is the clear signature of the hidden heteroclinic structure of the PR in the periodic NLSE. At variance with the well-known structure of the integrable focusing NLSE \cite{Akh85,Ablowitz89,Moon90,TW91}, it cannot be revealed directly from the space-time evolutions [Fig. \ref{f4}(a)], due to the fast scale oscillations associated with the driving.

\begin{figure}[h]
\centering
\includegraphics[width=\columnwidth]{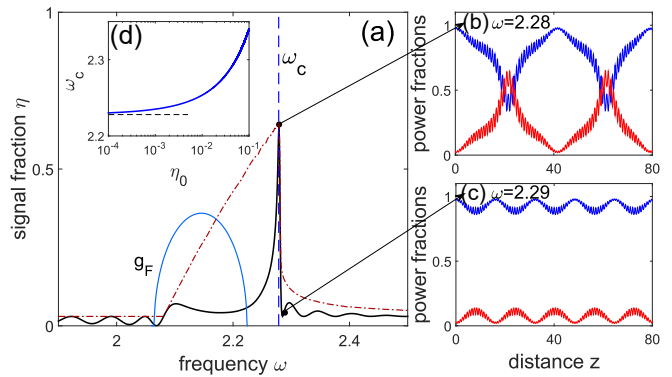}
\caption{(Color online) (a) Output sideband fraction $\eta(z=20)$ vs. $\omega$ from NLSE numerical integration for $\eta_0=0.03$ and $\phi_0=0$ (solid black curve; dash-dotted brown curve gives the maximum achievable conversion along $z$), with superimposed small-signal PR gain $g_F(\omega)$ (solid cyan curve). 
(b-c) Pump and sideband mode evolutions extracted from NLSE numerical integration across $\omega_c$ [vertical dashed line in (a), obtained from Eqs. (\ref{Hav})].  
Inset (d): $\omega_c$ vs. input modulation fraction $\eta_0$ (the horizontal dashed line stands for the edge frequency of $g_F(\omega)$).
     }
     \label{f5}
\end{figure}

The geometric structure of the nonlinear PR has even more striking consequence in terms of optimal parametric amplification of small sideband pairs. Clearly, the Floquet analysis entails that the sidebands growth rate is maximum at frequencies where the gain $g_F$ peaks. However, the nonlinear analysis shows that stronger conversion occurs towards higher frequencies of the gain curve, despite a slower initial growth. Indeed the long range conversion is associated  with quasi-periodic evolutions in the neighbourhood of the averaged separatrix, and the latter extends to larger portion of the phase space and hence larger values of $\eta$ as the frequency increases \cite{supplemental}. The remarkable and unexpected fact, however, is that strong nonlinear conversion occurs also at frequencies higher than the high-frequency edge of the PR bandwidth. While at such frequency the background is stable, strong nonlinear conversion is permitted nearby the heteroclinic orbit that emanates from the unstable eigenmodulations. As a result, the converted sideband fraction as a function of $\omega$ exhibits a maximum slightly below a critical frequency $\omega_c$ which lies off-resonance, i.e. outside the Floquet gain bandwidth of PR, as shown in Fig. \ref{f5}(a). Across $\omega_c$ the conversion abruptly drops, as entailed by qualitatively different conversion regimes [Fig. \ref{f5}(c,d)] and remain low for $\omega>\omega_c$. The critical frequency corresponds to the evolution along the heteroclinic orbit in Fig. \ref{f2}(c) and can be calculated as the implicit solution of the equation $H_p(\eta_{s}(\omega), \phi=\pm \pi/2)=H_p(\eta_0, \phi_0)$. $\omega_c$ tends to the high-frequency edge of the Floquet bandwidth in the limit of vanishing input signal  $\eta_0 \rightarrow 0$, and substantially deviates from it when increasing $\eta_0$, even moderately, e.g. up to 10\%, as shown in Fig. \ref{f5}(d).

In summary, we have unveiled the underlying phase-space structure of PR in the defocusing NLSE.
This allowed us to reveal the existence of a PR breather solution dividing different and unexpected recurrent regimes. 
Moreover, this study establishes the intrinsic inadequacy of the linearised Floquet analysis to determine the frequency for optimal parametric amplification.

\begin{acknowledgments}
The present research was supported by IRCICA (USR 3380 CNRS), by the Agence Nationale de la Recherche in the framework of the Labex CEMPI (ANR-11-LABX-0007-01), Equipex FLUX (ANR-11-EQPX-0017), by the projects NoAWE (ANR-14-ACHN-0014), TOPWAVE (ANR-13-JS04-0004), and FOPAFE (ANR-12-JS09-0005), and by the Fonds Europ\'{e}en de D\'{e}veloppement Economique R\'{e}gional. 
S.T. acknowledges also the grant PRIN 2012BFNWZ2.
\end{acknowledgments}


\begin{thebibliography}{99}
\bibitem{norm} The field $\psi$ can be rescaled to give $P=1$. Although we set $P=1$ in the figures, we leave $P$ as a parameter in the equations in order to highlight the role of nonlinear terms.
\end{thebibliography}
\end{document}